\documentclass[aps,onecolumn,prc,10pt,showpacs,preprintnumbers,
               nofootinbib,float,superscriptaddress,longbibliography]{revtex4-1}
\usepackage{geometry}
\usepackage{epsfig}
\usepackage{color}
\usepackage{float,amsmath,amssymb,diagbox}
\usepackage{multirow}
\usepackage{subfigure}
\usepackage{graphicx}
\usepackage{url}
\usepackage[utf8]{inputenc}
\usepackage{hyperref}
\usepackage{lineno}
\RequirePackage{graphicx}
\usepackage{slashed}
\usepackage{cleveref}

\begin{document}

\title{\LARGE Collision Energy Dependence of the Breit-Wheeler Process in Heavy-Ion Collisions and its Application to Nuclear Charge Radius Measurements}
\author{Xiaofeng Wang}
\address{Key Laboratory of Particle Physics and Particle Irradiation (MoE), Institute of Frontier and Interdisciplinary Science, Shandong University, Qingdao, China}
\author{James Daniel Brandenburg}
\author{Lijuan Ruan}
\address{Physics Department, Brookhaven National Laboratory, New York, USA}
\author{Fenglan Shao}
\address{School of Physics and Physical Engineering, Qufu Normal University, Shandong, China}
\author{Zhangbu Xu}
\address{Physics Department, Brookhaven National Laboratory, New York, USA}
\author{Chi Yang}
\address{Key Laboratory of Particle Physics and Particle Irradiation (MoE), Institute of Frontier and Interdisciplinary Science, Shandong University, Qingdao, China}
\author{Wangmei Zha}
\address{School of Physical Sciences, University of Science and Technology of China, Hefei, China}


\date{\today}

\begin{abstract}
The collision energy dependence of the cross section and the transverse momentum distribution of dielectrons from the Breit-Wheeler process in heavy-ion collisions are computed in the lowest-order QED and found to be sensitive to the nuclear charge distribution and the infrared-divergence of the ultra-Lorentz boosted Coulomb field.
Within a given experimental kinematic acceptance, the cross section is found to increase while the pair transverse momentum ($\sqrt{\langle p_{T}^{2} \rangle}$) decreases with increasing beam energy. 
We demonstrate that the transverse-momentum component of Weizsäcker-Williams photons is due to the finite extent of the charge source and electric field component in the longitudinal direction. 
We further clarify the connection between the nuclear charge distribution and the kinematics of produced $e^+e^-$ from the Breit-Wheeler process, and propose a criterion for the validity of the Breit-Wheeler process in relativistic heavy-ion collisions.
Following this approach we demonstrate that the experimental measurements of the Breit-Wheeler process in ultra-relativistic heavy-ion collisions can be used to quantitatively constrain the nuclear charge radius. The extracted parameters show sensitivity to the impact parameter dependence, and can be used to study the initial-state and final-state effects in hadronic interactions. 
\end{abstract}

\maketitle

\section{Introduction}
\label{intro}
In 1934, Breit and Wheeler studied the process of the collision of two light quanta to create electron and positron pairs. 
At that time Breit and Wheeler also noted that it is hopeless to observe the pair formation in laboratory experiments with two beams of x-rays or $\gamma$-rays meeting each other due to the smallness of the cross section and insufficiently large available densities of photon quanta~\cite{PhysRev.46.1087}. 
In high-energy heavy-ion collisions, ultra-strong electromagnetic fields can be obtained from the Lorentz-contraction of highly charged nuclei~\cite{Bzdak:2011yy,Voronyuk:2011jd,Deng:2012pc,Roy:2015coa,PhysRevC.104.034907}, and produce observable physics outcomes. 
In a specific phase space, these intense electromagnetic fields can be quantized as a flux of quasi-real photons (Equivalent Photon Approximation, EPA)~\cite{vonWeizsacker:1934nji,Williams:1934ad}, providing a viable source of photons to achieve the Breit-Wheeler process in the laboratory.

Traditionally these photon-photon processes were expected to exist only in Ultra-Peripheral Collisions (UPCs)~\cite{ALICE:2013wjo,Bertulani:1987tz,Baur:2007zz,STAR:2004bzo} for which the impact parameter between the colliding nuclei is larger than twice the nuclear radius such that no nuclear overlap occurs. 
However, it has recently been realized that even in events with nuclear overlap, the dielectron production at very low transverse momentum originates from two photon interactions~\cite{STAR:2018ldd,ATLAS:2018pfw,Lehner:2019amb}.
The ability to measure the Breit-Wheeler process in events with nuclear overlap provides additional avenues to study the collision energy dependence of the Breit-Wheeler process.
%


In high-energy $e^{+}e^{-}$ collisions, the photons are assumed to be emitted from the electron or positron. 
As shown by reference~\cite{ParticleDataGroup:2020ssz} (Eq. 50.44 and Eq. 50.45), photon flux diverges at both high and low four-momentum transfer, therefore, it requires the cutoffs at the minimum and maximum virtuality with a finite four-momentum transfer for photons to be emitted from the electron or positron. 
Unlike in the case for $e^{+}e^{-}$ collisions, the photon flux doesn't diverge in heavy-ion collisions~\cite{Krauss:1997vr}, because the low transverse momentum photon flux is regulated by the finite Lorentz factor of the ion, and the high transverse momentum photon flux is naturally cut off by the finite electric field strength due to the finite size of ion's continuous charge distribution~\cite{brandenburg2021mapping}. The consequential difference is that  
in $e^+e^-\rightarrow e^+e^- e^+e^-$ process, the photons act as mediators with its energy, transverse momentum distribution (TMD) and helicity state determined by the scattering states of the initial and final electron in action~\cite{abbiendi_total_2000,Brandenburg:2022tna} while in heavy-ion UPC Breit-Wheeler process, the photons are predetermined from the external field as linearly polarized photons with their transverse momentum and energy spectra constrained by the ion geometry and Lorentz-boost factor.  
Therefore the photon source from heavy-ion collisions is crucial for the discovery of the Breit-Wheeler process and the investigation of the photon space-momentum-spin correlation (Wigner function). These QED properties can be further tested by the collision energy dependence of pair root mean square of transverse momentum ($\sqrt{\langle p_{T}^{2} \rangle}$) and cross section for photon-photon process in heavy-ion collisions. 


Furthermore, the STAR collaboration at RHIC~\cite{STAR:2018ldd} and the ATLAS collaboration at LHC~\cite{ATLAS:2018pfw} have found a significant $p_{T}$ broadening effect for the lepton pairs from photon-photon processes in hadronic heavy-ion collisions (HHICs) in comparison to those in UPCs. And they are explained by introducing the effect of either the Lorentz force from a trapped electromagnetic field~\cite{STAR:2018ldd} or Coulomb scattering~\cite{ATLAS:2018pfw} in the Quark-gluon plasma (QGP) created in the HHICs. However, recent measurements by CMS to control the impact parameters in the UPC without the influence of the thermal medium show that the photon photon process is dependent on the impact parameters~\cite{CMS:2020skx}.
This $p_{T}$ broadening effect has successfully been described by the generalized EPA (gEPA), lowest order QED, and Wigner function formalism, each of which include the impact parameter dependence~\cite{brandenburg2021mapping,Zha:2018tlq,STAR:2019wlg,Klusek-Gawenda:2020eja}, which illustrates the importance of considering the spatial distribution of the electromagnetic fields. 
Strong electromagnetic fields arising from the Lorentz-contraction of highly charged nuclei generate a large flux of high-energy quasi-real photons. 
It has been argued in many publications that the characteristic momentum for photons from the electromagnetic fields of a given nucleus is $\langle k_{\bot}^{2} \rangle$ $\propto$ $1/R^{2}$~\cite{Baltz:2007kq,STAR:2004bzo,kleinSTARlightMonteCarlo2017b,Klein:2020jom,Bertulani:1987tz} based on the uncertainty principle, where $R$ is the nucleus charge radius. 
In this article we clarify the connections among the photon transverse momentum, the pair transverse momentum in the Breit-Wheeler process, and the nuclear geometry, in order to demonstrate the procedure for using experimental results to constrain the charge radius of large nuclei~\cite{brandenburg2021mapping}.


This Article is structured as follows: in~\Cref{sec:QED}, we derive a general form of the cross section in the lowest-order QED; in~\Cref{TMD}, we discuss the connections among the transverse momentum distributions of photons, of the $e^+e^-$ pair from the Breit-Wheeler process, and the nuclear geometry; in~\Cref{sec:Virtuality}, we discuss the photon virtuality and present a criterion for the Breit-Wheeler process in heavy-ion collisions; in~\Cref{sec:numerical results}, we present numerical estimations for the collision energy dependence of the cross section and $\sqrt{\langle p_{T}^{2} \rangle}$ in peripheral and ultra-peripheral heavy-ion collisions. An example of the constraining power on the nuclear charge distribution is shown in~\Cref{sec:discussions}. Finally, the Article is summarized in~\Cref{sec:conclusions}.

\section{Lowest Order QED}
\label{sec:QED}

The pair creation in lowest-order two photon interaction can be depicted as a process with two Feynman diagrams contributing, as shown in Fig.2 of Ref.~\cite{Hencken:1994my}. There is an approximation commonly used for describing events: that of external fields generated by nuclei that are undeflected by the collision and travel along straight-line trajectories. 
Following the derivation of Ref.~\cite{Hencken:1994my,Alscher:1996mja}, the cross section for pair production of leptons is given by
\begin{equation}
\label{equation_cross_section}
    \sigma = \int d^{2}b d^{6}P(\vec{b}) = \int d^{2}q d^{6}P(\vec{q}) \int d^{2}b e^{i {\vec{q}} \cdot  {\vec{b}}},
\end{equation}
and the differential probability $d^{6}P(\vec{q})$ in QED at the lowest order is
\begin{align}
    \label{equation_differential_probability}
    \begin{split}
        d^{6}P(\vec{q}) & = (Z\alpha_{em})^{4} \frac{4}{\beta^{2}} \frac{d^{3}p_{+}d^{3}p_{-}}{(2\pi)^{6}2\epsilon_{+}2\epsilon_{-}} \\ 
        & \times \int d^{2}q_{1} \frac{F(N_{0})F(N_{1})F(N_{3})F(N_{4})}{N_{0}N_{1}N_{3}N_{4}} \\
        & \times {\rm{Tr}}\bigg\{(\slashed{p}_{-}+m)\bigg[N_{2D}^{-1}\slashed{u}_{1} (\slashed{p}_{-} - \slashed{q}_{1} + m)\slashed{u}_{2} \\
        & + N_{2X}^{-1}\slashed{u}_{2}(\slashed{q}_{1} - \slashed{p}_{+} +m)\slashed{u}_{1}\bigg] (\slashed{p}_{+}-m) \\
        & \times \bigg[N_{5D}^{-1}\slashed{u}_{2} (\slashed{p}_{-} - \slashed{q}_{1} - \slashed{q} + m)\slashed{u}_{1} \\ 
        & + N_{5X}^{-1} \slashed{u}_{1} (\slashed{q}_{1} + \slashed{q} - \slashed{p}_{+} + m)\slashed{u}_{2}\bigg] \bigg\},
    \end{split}
\end{align}
with
\begin{align}
    \label{equation_N}
    \begin{split}
        N_{0} & = -q_{1}^{2},  \\
        N_{1} & = -[q_{1} - (p_{+}+p_{-})]^{2},\\
        N_{3} & = -(q_{1}+q)^{2}, \\
        N_{4} & = -[q+(q_{1} - p_{+} - p_{-})]^{2}, \\
        N_{2D} & = -(q_{1} - p_{-})^{2} + m^{2},\\
        N_{2X} & = -(q_{1} - p_{+})^{2} + m^{2}, \\
        N_{5D} & = -(q_{1} + q - p_{-})^{2} + m^{2},\\
        N_{5X} & = -(q_{1} + q  - p_{+})^{2} + m^{2},
    \end{split}
\end{align}
where $b$ is the impact parameter, $p_{+}$ and $p_{-}$ are the momenta of the created leptons, $Z$ is nuclear charge number, $\alpha_{em}$ is fine structure constant, $\beta=v/c$ with $v$ being the velocity of the nucleus, $c$ is the speed of light in vacuum, $\epsilon_{+}$ and $\epsilon_{-}$ are the energies of the produced leptons, $F(N_{0})$ is the nuclear electromagnetic form factor, $m$ is the mass of the lepton, $u_{1,2}$ is the four-velocity divided by Lorentz contraction factor ($\gamma$) of ions 1 and 2, $q_{1,2}$ is the four momentum of the photon emitted by ions 1 and 2, the longitudinal components of $q_{1}$ are given by $q_{10} = \frac{1}{2}[(\epsilon_{+} + \epsilon_{-}) + \beta(p_{+z}+p_{-z})]$, $q_{1z} = q_{10}/ \beta$, $q = q_{2} - q_{1}$. 
In order to compute results at all impact parameters, where in general no simple analytical form is available, the multi-dimensional integration is performed with the VEGAS Monte Carlo integration routine~\cite{peterlepageNewAlgorithmAdaptive1978}. 

The nuclear electromagnetic form factor can be obtained via the Fourier transform of the charge distribution as
    \begin{equation}
    \label{equation_formFactor}
    F(k^{2}) = \int d^{3}re^{ik.r}\rho_{A}(r).
    \end{equation}

In this Article, we assume that the charges in the target and projectile nuclei are distributed according to the Woods-Saxon distribution ~\cite{Woods:1954zz} without any fluctuations or point-like structure as
    \begin{equation}
    \label{equation_charge_density}
    \rho_{A}(r)=\frac{\rho^{0}}{1+\exp[(r-R)/d]}
    \end{equation}
where the radius $R$ (Au: 6.38 fm) and skin depth $d$ (Au: 0.535 fm) are based on fits to low energy electron scattering data such that all deformations are assumed to be higher order and are ignored~\cite{DEVRIES1987495}, and $\rho^{0}$ is the density at the center of nucleus. The Fourier transform of the Woods-Saxon distribution does not have an analytic form, it was computed numerically for the following calculations.  

The EPA is used when deriving the cross section for pair production in Eq.~\eqref{equation_cross_section}. According to the EPA, the number spectrum of photons with energy $\omega$~\cite{Krauss:1997vr} manifest by the field of a single nucleus is:
    \begin{equation}
    \label{equation_photon_density}
    n(\omega) = \frac{(Ze)^{2}}{\pi\omega}\int_{0}^{\infty}\frac{d^{2}k_{\perp}}{(2\pi)^{2}}\left[\frac{F\left(\left(\frac{\omega}{\gamma}\right)^{2}+\overrightarrow{k}_{\perp}^{2}\right)}{\left(\frac{\omega}{\gamma}\right)^{2}+\overrightarrow{k}_{\perp}^{2}}\right]^{2}\overrightarrow{k}_{\perp}^{2},
    \end{equation}
where $\overrightarrow{k}_{\perp}$ is the photon transverse momentum, and $F\left(\left(\frac{\omega}{\gamma}\right)^{2}+\overrightarrow{k}_{\perp}^{2}\right)$ is the nuclear electromagnetic form factor.

\section{Transverse Momentum Distribution}
\label{TMD}
From Eq.~\eqref{equation_photon_density}, the photon density increases dramatically as $k_{\perp} \to 0$ and would diverge if it were not regulated by the $\omega/\gamma$ factor. 
This implies that the measurements of collision energy dependence of the $l^{+}l^{-}$ pair differential cross section and mean transverse momentum would be sensitive to the infrared-divergence term as evident from those equations. 
Specifically in  Eq.~\eqref{equation_photon_density}, the transverse momentum would be expected to increase with decreasing beam energy ($\gamma$) for the same kinematic acceptance of $e^+$ and $e^-$ with fixed $\omega$. 
Although it is commonly believed that the transverse momentum distribution of photons is due to uncertainty principle and therefore $k_{\perp}\propto 1/R$, we could demonstrate how to obtain photon transverse momentum in classic electromagnetism. At a given ultra-relativistic Lorentz boost ($\gamma$), the classical electric field from a charged nucleus can be expressed as 
    \begin{equation}
    \label{equation_electric_field}
    \vec{E} = \frac{Ze}{4\pi\varepsilon_{0}\gamma^{2}r^{2}(1-\beta^{2}\sin^{2}\theta)^{\frac{3}{2}}}\hat{r},
    \end{equation}
where $\varepsilon_{0}$ is the vacuum permittivity, $r$ is the distance from the center of the nuclear to field point, $\hat{r}$ is the direction from the center of nuclear to field point. and $\theta$ is the angle between the electric field line and the beam direction. 
For any finite $\beta$ and $\theta$, there is a small component of the electric field in the beam direction.
The magnetic field expression can be obtained from the electric field as
    \begin{equation}
    \label{equation_magnetic_field}
    \vec{B} = \frac{1}{c^{2}}(\vec{v}\times\vec{E}).
    \end{equation}
From Eq.~\eqref{equation_magnetic_field} it can be found that the magnetic field exists only in the transverse plane. 
Therefore, the propagation of the electromagnetic wave ($\vec{E} \times \vec{B}$) has a small but finite component in the radial direction on the transverse plane. 
The photon density is related to the energy flux of the electromagnetic fields~\cite{Vidovic:1992ik} $n(\omega) \propto \vec{S} = \frac{1}{\mu_{0}} \vec{E} \times \vec{B}$, where $\mu_{0}$ is vacuum permeability, and $\vec{S}$ is the Poynting vector. 
The transverse component of the photon momentum can be obtained by projecting the electric field along the beam direction in Eq.~\eqref{equation_electric_field} as $E_{\parallel}=E\cos{\theta}$ and integrating over the polar angle $\theta$: 
    \begin{equation}
    \label{equation_omegaovergamma}
    \frac{k_{\perp}}{\omega} = \frac{E_{\parallel}}{E} = \frac{1}{\gamma}.
    \end{equation}
This relationship clearly shows that the transverse component of photon momentum is due to the finite projection of electric field along the z-axis and is not directly related to the transverse size of the charge distribution. 
One can also understand this intuitively that in a cylindrically symmetric charge distribution with infinity extension along beam direction (z-axis), the electric field is strictly perpendicular to the z-axis and therefore the photons propagate strictly along z direction with no transverse momentum regardless of the transverse radius. 
Similarly, in the high-power laser-driven nonlinear Breit-Wheeler process~\cite{SLACPhysRevLett.79.1626}, the photon generated by the electron-laser collisions serves as an intermediate propagator and its divergence is cutoff by the finite duration of the laser pulse~\cite{SLACTridentQEDPRL2010} with a laser pulse length about 10 times that of the laser photon wavelength.

How then are the measurements of the Breit-Wheeler process sensitive to the nuclear geometry? 
Unlike in the case for an $e^+e^-$ collider, the photon flux does not diverge in UPCs because the low-virtuality photon flux is regulated by the finite Lorentz factor of the ions ($k^2 \ge (\omega/\gamma)^2~ \gtrsim ~(2$ MeV$)^2$, for STAR acceptance and collision energy) and the high-virtuality photon flux is naturally cut off by the finite field strength due to the finite size of the ion's charge distribution in the form factor ($k^2~\lesssim~(1/R)^2 \simeq (30$ MeV$)^2$)(e.g. Eq. (38)$-$(45) in Ref.~\cite{Vidovic:1992ik}). 
However, there is an additional important factor which makes the Breit-Wheeler process sensitive to the nuclear geometry. 
The WW photons are linearly polarized, and the two Feynman diagrams~\cite{Hencken:1994my} in Eq.~\eqref{equation_differential_probability} cancel at low $k_{\perp}$. 
The phase modulation is of the form $\exp{(-i\vec{b}\cdot\vec{k_{\perp}})}$, and depends on the impact parameter which is related to the nuclear geometry. 
This is also what results in an impact-parameter dependence of the Breit-Wheeler process. 
We note that the two-diagram interference dependence is usually absent in models~\cite{kleinSTARlightMonteCarlo2017b} implementing the Breit-Wheeler process at the cross section level and not at the quantum wavefunction level. 
Therefore, in high-energy ultra-peripheral heavy-ion collisions, the low-$k_{\perp}$ is modulated by $\exp{(-i\vec{b}\cdot\vec{k_{\perp}})}$
and high-$k_{\perp}$ by the form factor shown in Eq.~\eqref{equation_formFactor}. Both of these factors are a function of the nuclear geometry. 

\section{Photon Virtuality and a Criterion for the Breit-Wheeler Process}
\label{sec:Virtuality}
It is often considered that the transverse momentum of the photons in UPC is related to the transverse dimensions of the nuclei and the virtuality of the photons as discussed in the previous section. 
This has been used as an argument 
that the $e^+e^-$ pair production from UPC is not the Breit-Wheeler process despite the original proposal in the Breit-Wheeler paper~\cite{PhysRev.46.1087}. In this section, we follow the Vidovic paper~\cite{Vidovic:1992ik} using the S-Matrix derivation to illustrate the approximation which results in the EPA and propose a criterion for defining the Breit-Wheeler process in relativistic heavy-ion collisions. 

\begin{figure}
\centering
    \begin{minipage}[t]{0.49\textwidth}
    \includegraphics[width=7.5cm]{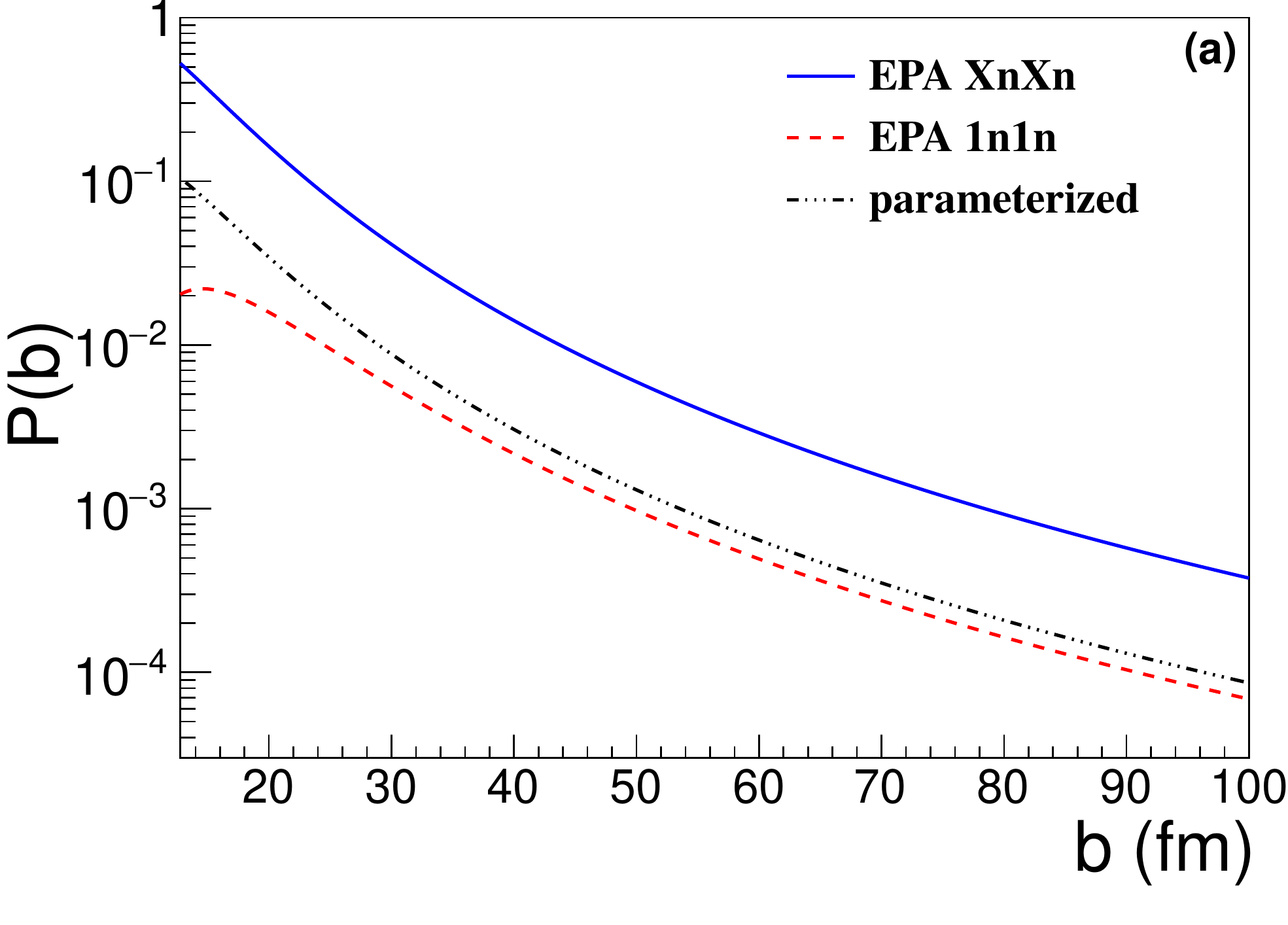}
    \end{minipage}
    \begin{minipage}[t]{0.49\textwidth}
    \includegraphics[width=7.5cm]{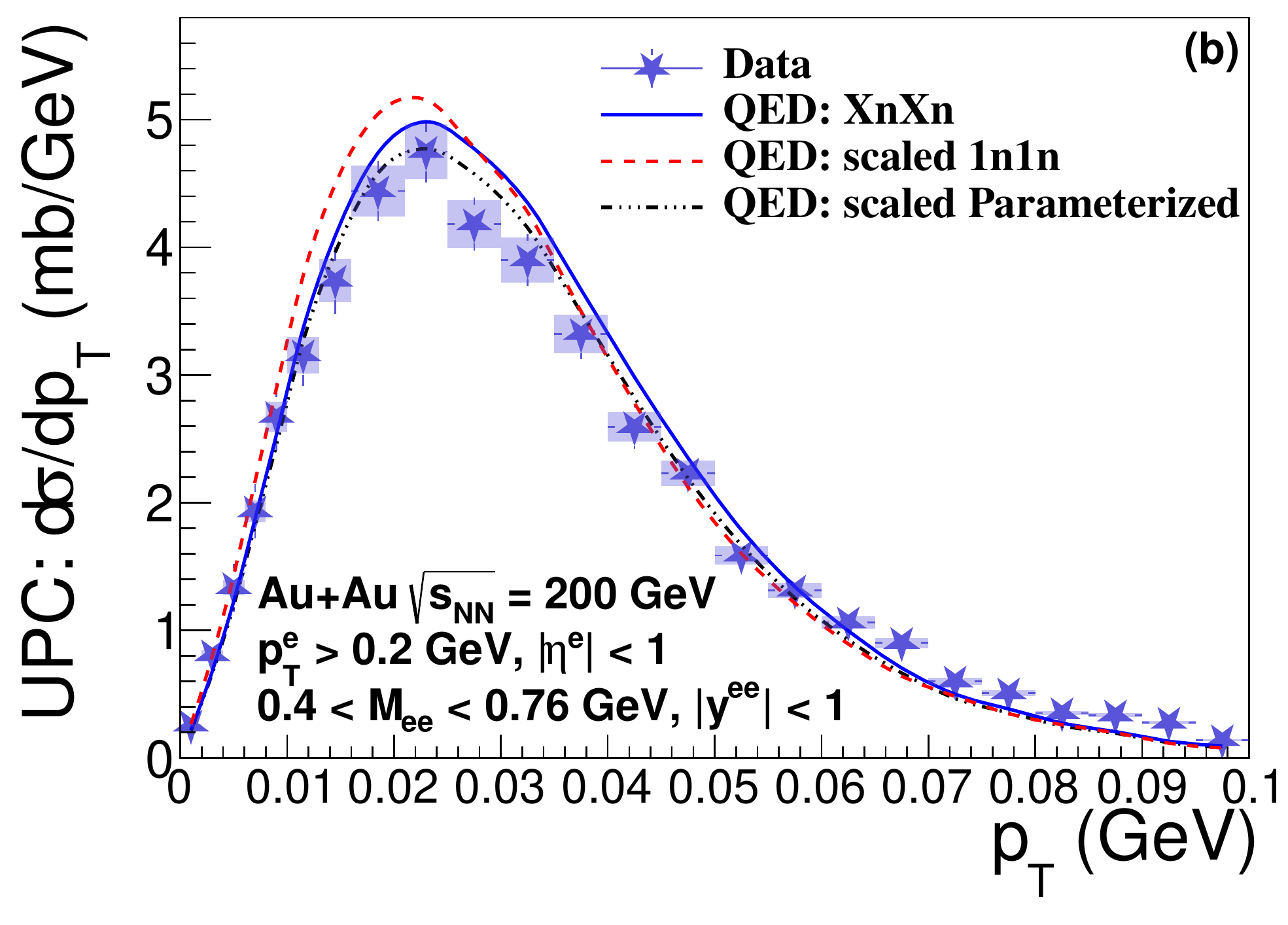}
    \end{minipage}
\caption{(color online) (a) The gold nucleus break-up probability as a function of impact parameter based on parameterized method and EPA method with different neutron selections. (b) Differential cross sections as a function of dielectron transverse momentum according to the probabilities shown in the left plot compared to the STAR measurement~\cite{STAR:2019wlg} with neutron selection condition XnXn in Au+Au UPCs at 200 GeV.}
\label{fig: probability of nuclear dissociation}
\end{figure}

\begin{figure}[htbp]
 \centering
\setlength{\abovecaptionskip}{0.cm}
    \includegraphics[width=0.5\textwidth]{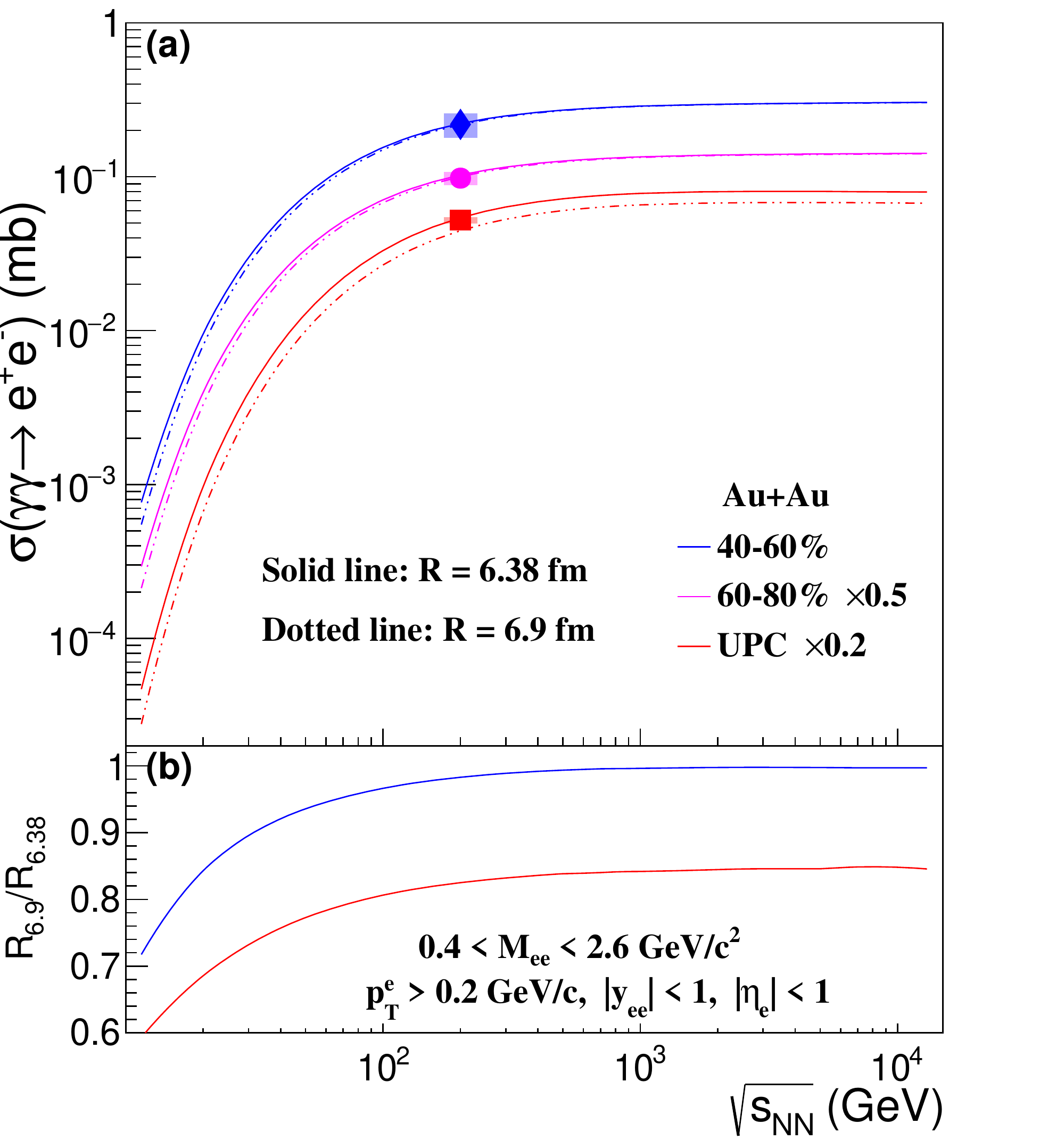}
    \caption{ (color online) (a) The cross section for the production of $e^+e^-$ pairs via the Breit-Wheeler process in Au+Au collisions within STAR acceptance as a function of center-of-mass energy. Results are shown for different centralities and for two different nuclear charge radii of 6.38 fm (solid line) and 6.9 fm (dotted line). The STAR measurements~\cite{STAR:2018ldd, STAR:2019wlg} are also plotted for comparison. (b) The corresponding ratios of the cross section for $R = 6.9$ fm over $R = 6.38$ fm.
}
    
    \label{fig:CrossSection}
\end{figure}

Since the Coulomb field is a pure electric field, the Lorentz boost does not change the fact that real photons cannot be generated by the single standalone nuclear field itself. The resulting quantization as photons from the Poynting Vector from one nucleus would have the form as shown in Eq.~\eqref{equation_photon_density} with a spacelike Lorentz vector and a "negative squared mass" of $-((\omega/\gamma)^2+k_{\perp}^2)$. 
It was argued that if one were to define the process in UPC as the Breit-Wheeler process, the virtuality would simply have to be ignored. This is not the case. In fact, setting this term to zero would result in infrared divergence of the photon flux. 
Equation 25 in Ref.~\cite{Vidovic:1992ik} shows the approximation required for the conserved current of the transition probability from $\gamma \gamma$ to $l^+l^-$ pairs in the S-Matrix to behave like real photon interactions. The requirement in the center-mass-frame of the heavy-ion collision is that both photons satisfy the following condition: 
    \begin{equation}
    \label{equation_BW_criterion1}
    \omega/\gamma\, \lesssim \, k_{\perp}\, <<\, \omega     \end{equation}

With this criterion and subsequent omission of the higher order second and third terms of the order of $1/\gamma^2$, the vertex function of the two-photon process in relativistic heavy-ion collisions in Eq. 28 of Ref.~\cite{Vidovic:1992ik} would be identical to that of the real-photon interaction in Eq. 19 of Ref.\cite{Vidovic:1992ik}. 
The interpretation is therefore that the single photon flux of the virtual states from the Lorentz boosted field is given by Eq.~\eqref{equation_photon_density} and that the interaction is only relevant for (or behaves as) photons with real-photon states characterized by energy of $\omega$ and transverse momentum of $k_{\perp}$, validating the implementation of the so-called photon Wigner function (PWF)~\cite{Klein:2020jom,Li:2019sin,Klusek-Gawenda:2020eja,Wang:2021kxm}. 
The form factor (field strength) in the photon flux limits the photon transverse momentum to be $k_{\perp} \lesssim 1/R$ and in the regime of much higher $k_{\perp}$ ($k_{\perp}\gtrsim1/R$ and/or $\omega\gtrsim\gamma/R$),  significant contributions from the "semi-coherent" process~\cite{Staig:2010by} with photon scattering off constituent nucleons and quarks inside nucleus may invalidate the EPA assumption. This puts a further constraint on the available phase space for the photons that may participate in the Breit-Wheeler process: 
    \begin{equation}
    \label{equation_BW_criterion2}
    \omega/\gamma\, \lesssim\, k_{\perp}\, \lesssim\, 1/R\, \ll\, \omega     \end{equation}
With decreasing beam energy ($\gamma$) in the same kinematic acceptance, the phase space for the Breit-Wheeler process decreases and we would expect that the photons outside this valid range ($k_{\perp}\lesssim\omega/\gamma$) to contribute substantially to the interaction cross section at low beam energy. 

\begin{figure}[htb]
  \begin{center}
    \includegraphics[width=0.5\textwidth]{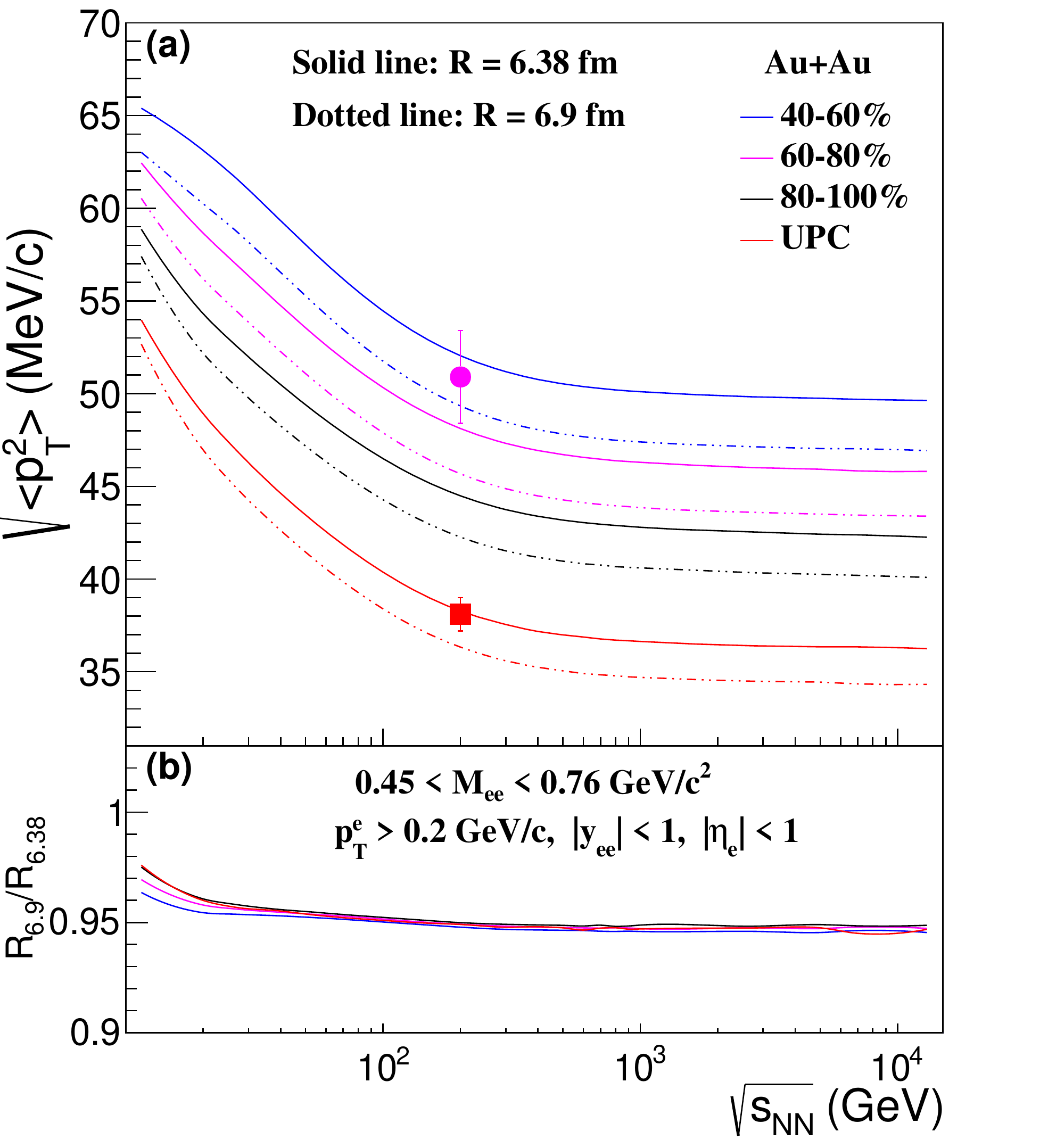}
  \end{center}    
  \caption{ \label{fig:MeanPt} (color online) (a) The $\sqrt{\langle p_{T}^{2} \rangle}$ of $e^{+}e^{-}$ pairs produced in Au + Au collisions within STAR acceptance as a function of center-of-mass energy. Results are shown for different centrality and for nuclear radii of 6.38 fm (solid line) and 6.9 fm (dotted line). The STAR measurements~\cite{STAR:2018ldd, STAR:2019wlg} are also plotted for comparison. (b) The corresponding cross section ratios for $R = 6.9$ fm over $R = 6.38$ fm.}
\end{figure}

\section{Numerical Results}
\label{sec:numerical results}

In this paper, we focus on peripheral and ultra-peripheral collisions. In peripheral collisions, the Breit-Wheeler process may be accompanied by hadronic interactions. According to the optical Glauber model, the mean number of projectile nucleons that interact at least once in an A+A collision with impact parameter $b$ is~\cite{Miller:2007ri,Brandenburg:2020ozx}:
\begin{equation}
N_{H}(b) = \int d^{2}\vec{r} T_{A}(\vec{r} - \vec{b}) \{ 1 - exp[-\sigma_{\rm{NN}}T_{A}(\vec{r})]\},
\end{equation}
with the nuclear thickness function ($T_{A}(\vec{r})$) determined from the nuclear density distribution:
\begin{equation}
T_{A}(\vec{r}) = \int dz \rho(\vec{r},z),
\label{equation_ThicknessFunction}
\end{equation}
where $\sigma_{\rm{NN}}$ is the total nucleon-nucleon inelastic cross section, and the subscript $H$ of $N_{H}$ stands for hadronic collisions. The collision energy dependence of $\sigma_{\rm{NN}}$ has been determined via a fit utilizing the parameterization $\sigma_{\rm{NN}}(s) = A + B ln^{n}(s)$~\cite{PhysRevC.97.054910}. In this work, we use values of $A = 25.0$ mb, $B = 0.146$ mb, $s$ in the unit of $(GeV)^2$ and $n = 2$ in numerical calculation.  Then, the probability of having a hadronic interaction ($1 - exp[-N_{H}(b)]$) can be obtained, which is also used to determine the collision centrality. 

\begin{figure}
\centering
    \begin{minipage}[t]{0.49\textwidth}
    \includegraphics[width=7.5cm]{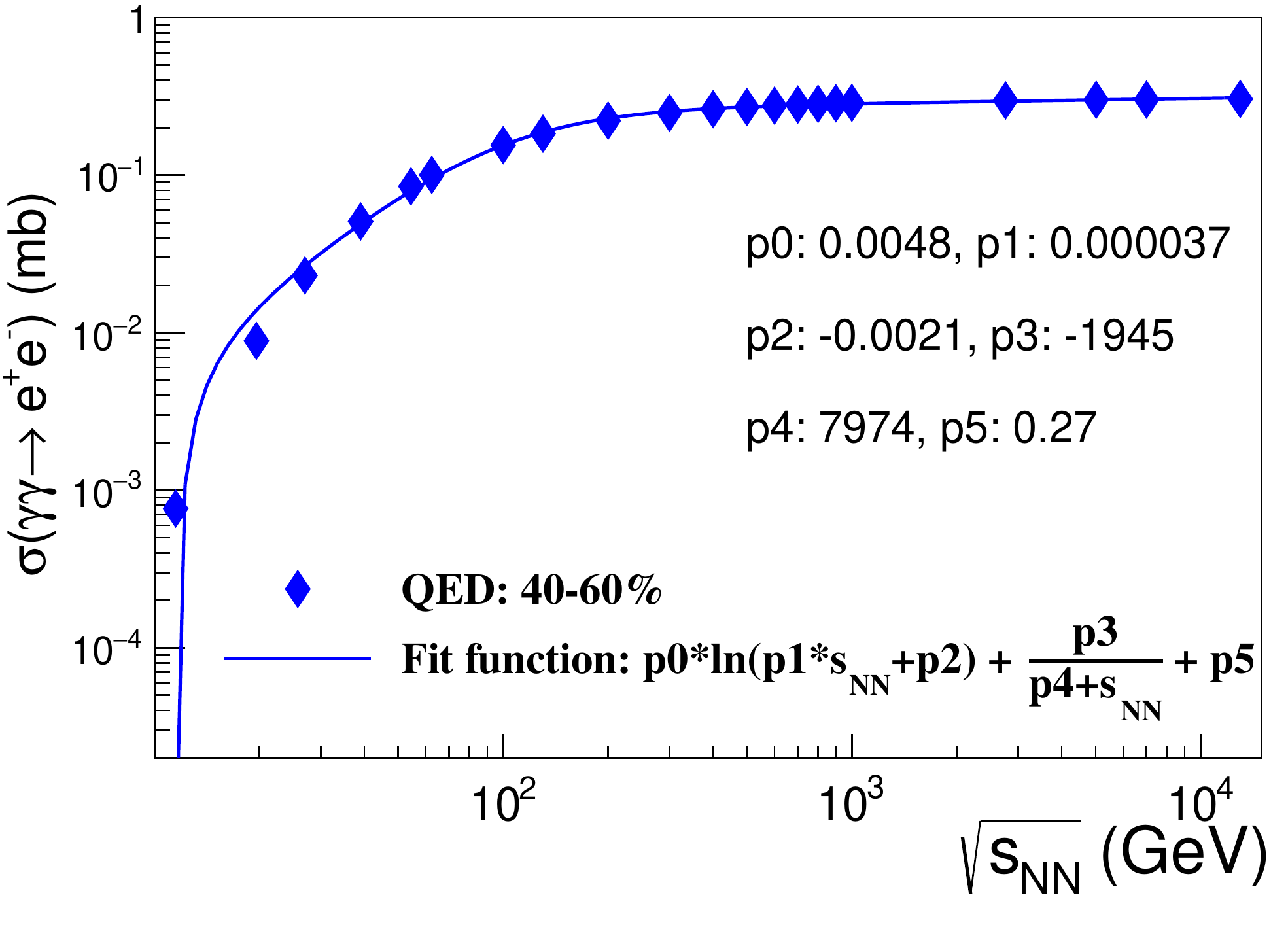}
    \end{minipage}
    \begin{minipage}[t]{0.49\textwidth}
    \includegraphics[width=7.5cm]{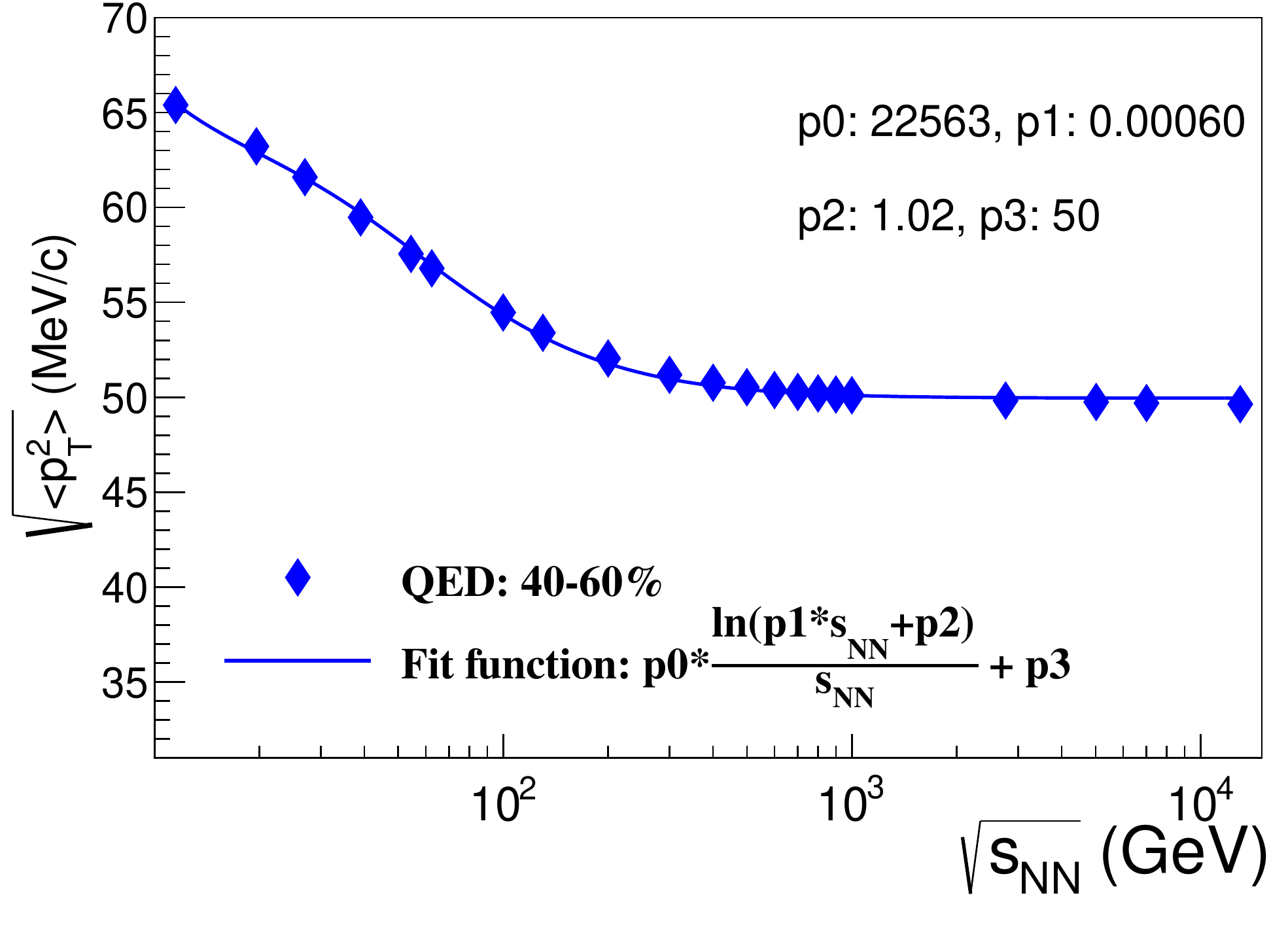}
    \end{minipage}
\caption{(color online) Cross section (left panel) and $\sqrt{\langle p_{T}^{2} \rangle}$ (right panel) as a function of center-of-mass energy fitted by Eq.~\eqref{equation_cross_section1} and Eq.~\eqref{equation_mean_pt1} respectively.}
\label{fig: fit QED}
\end{figure}

For ultra-peripheral collisions, the nuclei pass one another with a nucleus-nucleus impact parameter $b$ large enough such that there are no hadronic interactions. 
So, for UPCs, the probability of having no hadronic interaction ($exp[-N_{H}(b)]$) must also be taken into account, especially for $b \sim 2R$. 
The density of photons provided by the fields of highly charged nuclei is appreciable, therefore, the nuclei may exchange multiple photons in a single passing, which lead to the excitation and subsequent dissociation of the nuclei. 
The STAR experiment at RHIC measures cross section of pair production together with the mutual electromagnetic excitation of the nuclei, in which neutrons are emitted from both ions. The neutron emission multiplicity can be selected in the zero-degree calorimeter (ZDC)~\cite{Xu:2016alq,STAR:2002eio}. In order to incorporate the experimental conditions into the theoretical calculations, the probability of emitting neutrons from an excited nucleus must be included. The 1n1n is defined as two colliding nuclei that each emit a neutron, while XnXn is defined as colliding nuclei that each emit at least one neutron. The probabilities of 1n1n and XnXn can be obtained using the EPA method~\cite{Brandenburg:2020ozx}. The parameterized $P(b)$ shown in Eq.~\eqref{equation_parameterized_probability} is another widely adopted method to describe the probability of emitting a neutron from the scattered nucleus~\cite{Li:2019sin,Wang:2021kxm}. Fig.~\ref{fig: probability of nuclear dissociation} (a) shows the probability distributions as a function of impact parameter based on parameterized method and EPA method with different neutron selections. Fig.~\ref{fig: probability of nuclear dissociation} (b) shows differential cross sections as a function of dielectron transverse momentum according to the probabilities shown in Fig.~\ref{fig: probability of nuclear dissociation} (a). The differential cross sections with probabilities of 1n1n by EPA method and parameterized method are scaled to compare to the published XnXn UPC data~\cite{STAR:2019wlg}. It is clear that the calculation with all three probabilities can reasonably describe the shape of measured $p_{T}$ distribution.

\begin{align}
\label{equation_parameterized_probability}
   \begin{split}
    P(b) & = 5.45 * 10^{-5} \frac{Z^{3}(A - Z)}{A^{2/3}b^{2}}\\
    & \times exp\bigg[- 5.45 * 10^{-5} \frac{Z^{3}(A - Z)}{A^{2/3}b^{2}} \bigg].
    \end{split}
\end{align}

\begin{figure}[htb]
  \begin{center}
    \includegraphics[width=1.\textwidth]{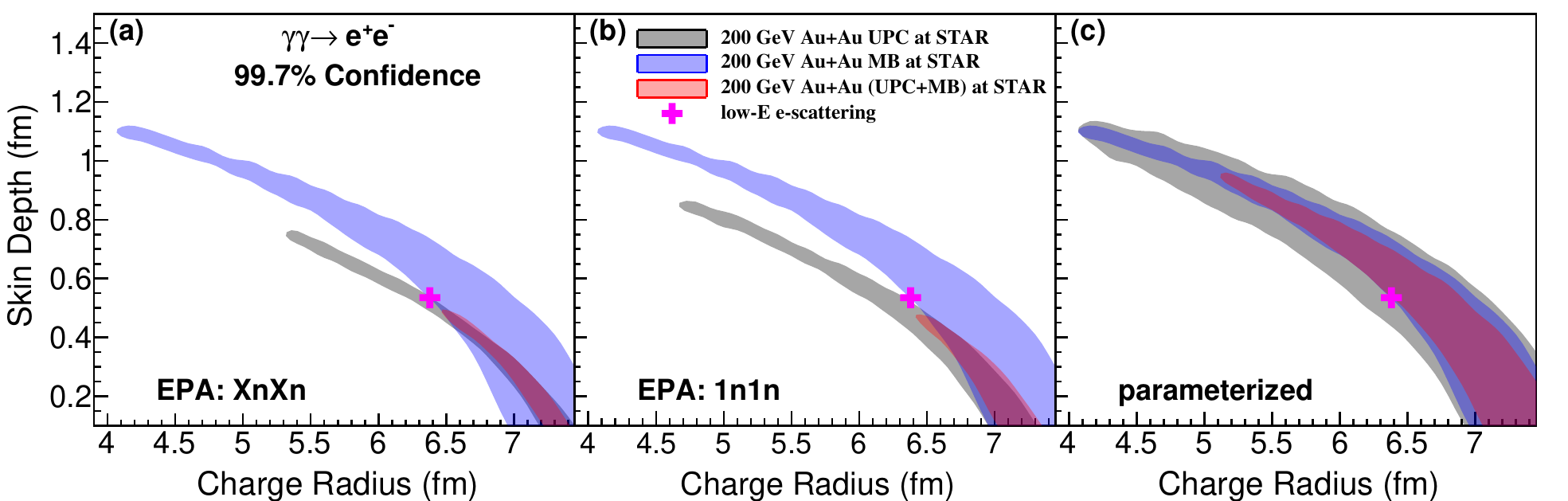}
  \end{center}    
  \caption{ \label{fig:Chi2} (color online) The constraints on gold nuclear charge distribution obtained by the comparison between STAR measurement of $\gamma\gamma \rightarrow e^{+}e^{-}$ and the lowest order QED calculation for different neutron selection conditions in ZDC and parameterized probability, (a), (b), and (c) are for XnXn, 1n1n, and parameterized probability, respectively.}
\end{figure}

We follow STAR experimental conditions in choosing to integrate the rapidities and transverse momentum of the electron (positron) over the ranges [-1, 1] and [0.2 GeV, 1.4 GeV], respectively. Similarly, the transverse momentum of the $e^{+}e^{-}$ pair is required to be less than 200 MeV. Neutron selection condition XnXn is used to get Fig.~\ref{fig:CrossSection} and Fig.~\ref{fig:MeanPt}.
We plot the cross section and $\sqrt{\langle p_{T}^{2}} \rangle$ of $e^{+}e^{-}$ pairs as a function of center-of-mass energy within STAR acceptance for peripheral and ultra-peripheral collisions in Fig.~\ref{fig:CrossSection}(a) and Fig.~\ref{fig:MeanPt} (a). The general trend is that the cross section increases while the $\sqrt{\langle p_{T}^{2}} \rangle$ decreases when the center-of-mass energy increases. Both the cross section and the $\sqrt{\langle p_{T}^{2}} \rangle$ tend to reach a plateau at higher energy for the same kinematic acceptance. 
As discussed earlier, and now numerically demonstrated, $\sqrt{\langle p_{T}^{2}} \rangle$ has a significant dependence on impact parameter and does not follow the photon $k_{\perp}$ decrease as $\omega/\gamma$ at high energy. 
The turning point where the plateau sets in is at beam energy of around 100 GeV, and therefore, the RHIC beam energy range of $\sim 20-200$ GeV is ideal for studying this effect with the generic detector capabilities available in the high-energy nuclear physics.  

To get some qualitative understanding of the trends of the beam energy dependence shown in Fig.~\ref{fig:CrossSection} and Fig.~\ref{fig:MeanPt} that we obtain from the numerical calculations, we derive the cross section and $\sqrt{\langle p_{T}^{2}} \rangle$ from the photon density in Eq.~\eqref{equation_photon_density} with further approximations of a Gaussian nuclear charge profile and additive photon momenta~\cite{kleinSTARlightMonteCarlo2017b}. The resulting equations for the cross section and $\sqrt{\langle p_{T}^{2}} \rangle$ are shown as Eq.~\eqref{equation_cross_section1} and Eq.~\eqref{equation_mean_pt1}. 

    \begin{equation}
    \label{equation_cross_section1}
    \sigma \propto ln(\frac{\gamma^{2}}{R^{2}}+\omega^{2}) + \frac{\omega^{2}R^{2}}{\omega^{2}R^{2}+\gamma^{2}} + C
    \end{equation}
    
    \begin{equation}
    \label{equation_mean_pt1}
    \sqrt{\langle p_{T}^{2} \rangle} \propto 2\frac{\omega^{2}}{\gamma^{2}}ln(\frac{\gamma^{2}}{R^{2}}+\omega^{2}) + C
    \end{equation}
Where $C$ is some integration constant.  These are used to fit the QED calculations for cross section and $\sqrt{\langle p_{T}^{2} \rangle}$ and illustrated for 40-60\% centrality as a function of beam energy in Fig.~\ref{fig: fit QED}. These logarithmic functions can describe the overall trends of our QED calculations.

\begin{table*}[htbp]
  \centering
  \caption{\label{table: RMS of radius in minimum chi2 plus 1}RMS of radius ($\sqrt{\langle r^2\rangle}$) at minimum $\chi^2$ ($\chi^{2}_{min}$) and uncertainties within $\chi^{2}_{min} + 1$ with different $\sigma_{_{NN}}$ and with different neutron selection conditions in ZDC and parameterized probability. UPC and MB respectively represent ultra-peripheral collision and minibias collision data are used, which is same as Fig.~\ref{fig:Chi2}. A default $\sigma_{_{NN}}=41.6$ mb has been used in all other calculations. These are to be compared to the default value of nuclear charge radius RMS of $\sqrt{\langle r^2\rangle}=5.33$ fm at $R=6.38$ fm and $d=0.535$ fm. }
  \setlength{\tabcolsep}{3.5mm}{
\begin{tabular}{|c|c|c|c|c|}
 \hline
 \textbf{condition} & \textbf{$\sigma_{_{NN}}$ (mb)} & \textbf{UPC}          & \textbf{MB}            & \textbf{UPC+MB} \\ \hline
 \multirow{4}{*}{\textbf{1n1n}} & 35.0 & 5.55 + 0.03 - 0.30  & 5.66 + 0.09 - 0.12   & 5.55 + 0.03 - 0.03\\
 & 40.0              & 5.32 + 0.26 - 0.21  & 5.67 + 0.08 - 0.10  & 5.58 + 0.01 - 0.04\\ 
 & 41.6    & 5.39 + 0.14 - 0.21  & 5.67 + 0.08 - 0.12   & 5.53 + 0.10 - 0.02\\ 
 & 45.0              & 5.47 + 0.02 - 0.21  & 5.66 + 0.09 - 0.11   & 5.54 + 0.08 - 0.03\\   \hline
 \multirow{4}{*}{\textbf{XnXn}} & 35.0 & 5.70 + 0.01 - 0.29  & 5.66 + 0.09 - 0.12   & 5.64 + 0.07 - 0.07\\
 & 40.0              & 5.70 + 0.01 - 0.30  & 5.67 + 0.08 - 0.10   & 5.70 + 0.01 - 0.12\\ 
 & 41.6    & 5.67 + 0.03 - 0.17  & 5.67 + 0.08 - 0.12   & 5.67 + 0.03 - 0.09\\ 
 & 45.0              & 5.54 + 0.17 - 0.16  & 5.66 + 0.09 - 0.11   & 5.64 + 0.06 - 0.11\\   \hline
 \multirow{4}{*}{\textbf{Parameterized}} & 35.0 & 5.51 + 0.15 - 0.18  & 5.66 + 0.09 - 0.12   & 5.61 + 0.13 - 0.11\\
 & 40.0              & 5.43 + 0.22 - 0.08  & 5.67 + 0.08 - 0.10   & 5.67 + 0.04 - 0.16\\ 
 & 41.6    & 5.41 + 0.25 - 0.09  & 5.67 + 0.08 - 0.12   & 5.62 + 0.12 - 0.11\\ 
 & 45.0              & 5.40 + 0.23 - 0.17  & 5.66 + 0.09 - 0.11   & 5.62 + 0.09 - 0.11\\   \hline
\end{tabular}}
\end{table*}

In order to investigate the dependence on the nuclear charge distribution, we vary the Woods-Saxon radius of the nuclear charge distribution in the calculations from $6.38$ fm to $6.9$ fm. The recalculated values of the cross section and $\sqrt{\langle p_{T}^{2} \rangle}$ with $R=6.9$ fm are shown as dotted lines in Fig.~\ref{fig:CrossSection}(a) and~\ref{fig:MeanPt} (a). The corresponding ratios of the cross section and $\sqrt{\langle p_{T}^{2} \rangle}$ for $R = 6.9$ fm over $R = 6.38$ fm are shown in Fig.~\ref{fig:CrossSection}(b) and~\ref{fig:MeanPt}(b), respectively.  These results show that the larger the radius of the nuclear charge distribution, the smaller the cross section and the smaller the average $e^+e^-$ pair momentum. The ratios which deviate from unity demonstrate that the kinematics of the $\gamma\gamma \rightarrow e^{+}e^{-}$ process are sensitive to the details of the nuclear charge distribution.

Figure~\ref{fig:Chi2} shows the $99.7\%$ ($3\sigma$) confidence level contour for the extracted nuclear charge distribution for a gold nucleus. These confidence contours result from a $\chi^{2}$-minimization procedure applied to the STAR measurements of the pair $p_{T}$ and invariant mass ($M_{ee}$) distributions from the $\gamma\gamma \rightarrow e^{+}e^{-}$ process~\cite{STAR:2018ldd,STAR:2019wlg} compared to the corresponding lowest-order QED calculations with the probabilities based on parameterized method and EPA method. For the minimization, the nuclear radius and skin depth are parameterized according to a Woods-Saxon distribution and are assumed to be the same for both electromagnetic and strong interactions. And the confidence level contours are all from experimental uncertainty only. The QED in UPC calculated with the parameterized probability has an arbitrary normalization and an additional overall normalization parameter is used to fit the data to get the $\chi^{2}$ while the absolute cross section is used to obtain the $\chi^{2}$ from the other two neutron emission distributions (1n1n and XnXn). The charge radius can be better constrained when the cross section is also taken into account in the $\chi^{2}$ calculations shown as the contours in Fig.~\ref{fig:Chi2} (a) and (b) compared to Fig.~\ref{fig:Chi2} (c). The results show that the RHIC measured charge radius deviates systematically from that from low-energy electron scattering at the $2-3\sigma$ level. Compared to Fig. 8 in Ref.~\cite{brandenburg2021mapping} obtained from only the measured $p_{T}$ distribution in UPCs, the gray contour in Fig.~\ref{fig:Chi2} obtained from the measured $p_{T}$ and $M_{ee}$ distributions in UPCs~\cite{STAR:2019wlg} shows a trend towards a slightly larger radius. In addition to adding the $M_{ee}$ distribution in constraining the nuclear parameters, another difference is that our current result uses the same free parameters of the form factors for both the strong-interaction radius and the charge radius. On the other hand, the result in Fig. 8 of Ref.~\cite{brandenburg2021mapping} was found with only the charge radius in the form factor as a free parameter, assuming that the strong-interaction radius of the nucleus was unchanged.

Another important factor we need to consider is the model uncertainties. There are a few model conditions which contribute to the numerical variations of the model implementation. The selection of neutron emission multiplicity in the ZDC provides an effective cut-off at large impact parameter. There are different implementations of such a probability distribution function obtained from experimental data and theory. The UPC data published by the STAR Collaboration are with 1-4n selections and are scaled up with an overall scale factor to match with XnXn. In reality, the impact parameter distribution is closer to an 1n1n selection~\cite{STAR:2019wlg}. \Cref{table: RMS of radius in minimum chi2 plus 1} lists the obtained root-mean-square (RMS) of the charge radius from the comparison between experimental data and model under different conditions. Specifically, the results from UPC show that the obtained radius is slightly larger (by about 0.1 fm) if XnXn neutron selection is assumed than if 1n1n selection is assumed. The other uncertainty is the nucleon-nucleon inelastic interaction cross section ($\sigma_{_{NN}}$). This condition limits how smaller the impact parameter could be in UPCs. \Cref{table: RMS of radius in minimum chi2 plus 1} also lists the obtained RMS of radius for different $\sigma_{_{NN}}$. The effect is at the level of 0.1 fm when $\sigma_{_{NN}}$ changes from 35 mb to 45 mb. We emphasize that these three model conditions only affect the impact-parameter probability distribution and do not change other terms in the calculations presented in the previous sections. The transverse momentum distribution has a finite variation as a function of impact parameter~\cite{Zha:2018tlq}, however, that variation in UPCs with the ZDC selection on the mutual Coulomb disassociation is small.  Similarly, the difference between a Glauber model with continuous density distribution function vs. that with a Monte Carlo simulation would be even smaller because those two different Glauber models would result in the same probability distribution. In the peripheral collisions, the $\sigma_{_{NN}}$ would change the space distribution of the participant nucleons but would not change the centrality definition and its impact parameter value.

\section{Discussions}
\label{sec:discussions}

\Cref{fig:CrossSection} shows a logarithmic growth of cross section and Fig.~\ref{fig:MeanPt} shows a flat distribution for $\sqrt{s_{_{NN}}}\geq100$ GeV ($\gamma\geq50$). These are consistent with the discussion in~\Cref{sec:Virtuality} of the Breit-Wheeler process. For lower energy ($\gamma\leq50$), the numeric results show that the cross section decreases dramatically while $\sqrt{\langle p_{T}^{2} \rangle}$ increases with decreasing beam energy. 
This is consistent with the substantial contributions of photon interaction from phase space with $k_{\perp}{}^<_{\sim}\omega/\gamma$. 
Within the kinematic acceptance, it was required that the single electron (positron) momentum to be $>200$ MeV at midrapidity~\cite{STAR:2018ldd,STAR:2019wlg}. 
This momentum threshold requires $\gamma\geq10$ to have any phase space for the Breit-Wheeler process as defined in Eq.~\eqref{equation_BW_criterion2}. The results shown in Fig.~\ref{fig:CrossSection} and Fig.~\ref{fig:MeanPt} suggest that significant contributions to the process outside of that valid range start at $\gamma\lesssim50$. These highlight the importance of relevant kinematics when we discuss the validity of the Breit-Wheeler process and its specific trend as a function of beam energy and centrality. 
Conversely, at extreme high energy, there are constraints on the validity of the Breit-Wheeler process as well. 
We note that in addition to the lepton pair momentum, the acoplanarity ($\alpha$) has been used in literature~\cite{ATLAS:2018pfw}. 
The criterion can be readily defined in terms of acoplanarity since it is straightforwardly related as  $\sqrt{2}k_{\perp}\simeq{\frac{\pi}{2}}\omega\alpha$~\cite{ATLAS:2022vbe, Zha:2018tlq}. Therefore, the criterion of the Breit-Wheeler process in terms of acoplanarity reads: 
    \begin{equation}
    \label{equation_BW_criterion3}
    {\frac{\sqrt{2}}{\gamma}} \lesssim{\frac{\pi}{2}}\alpha\lesssim{\frac{\sqrt{2}}{\omega R}}<<1    \end{equation}
For the kinematics of the ATLAS experiment at the LHC~\cite{ATLAS:2018pfw,ATLAS:2022vbe} with $\gamma=2500$ and $\omega \gtrsim 10$ GeV, the real-photon criterion becomes $4\lesssim k_{\perp}\lesssim30$ MeV (or $0.0004\lesssim\alpha\lesssim0.003$). 
Recent ATLAS results~\cite{ATLAS:2022vbe} of $p_{T}$ and $\alpha$ in central Pb+Pb collisions show that the full QED calculation presented in this article can describe the depletion at $\alpha\simeq0$ better than the PWF~\cite{Klein:2020jom}. 
We postulate that this difference may be due to the breakdown of the real-photon approximation in PWF at the extreme phase space when both photon $k_{\perp}$($\lesssim4$ MeV) approach zero and the conserved transition current could not be approximated as two real-photon vertex function as discussed in~\Cref{sec:Virtuality}, and the Landau-Lifshitz process for the collisions of two virtual photons may have to be considered~\cite{landauCreationElectronsPositrons1934}.  

In the approach described in the current draft, we ignore several potential effects at the initial and final stages. They are: (a.) initial charge fluctuation~\cite{Skokov:2009qp,Bzdak:2011yy,Miller:2007ri} and semi-coherent scattering~\cite{Staig:2010by,Miller:2007ri}; (b.) higher order Coulomb correction and suppression~\cite{Sun:2020ygb,Baltz:2007gs,lee2009strong,zha2021discovery,Hencken:2006ir}; (c.) final-state Sudakov radiation~\cite{Klein:2020jom,Brandenburg:2022jgr}; (d.) electromagnetic interaction with the surrounding medium~\cite{STAR:2018ldd,ATLAS:2018pfw,Klein:2020jom,BURMASOV2022108388}. Each effect has specific observable features in addition to the effects on the broadening of transverse momentum and the suppression of the overall cross sections. Until those specific features are observed, it is difficult to take those effects into account when the lowest-order QED has been demonstrated to be able to describe the existing data to high precision. It is beyond the scope this paper and we refer readers to a few recent review papers on these topics~\cite{Brandenburg:2022tna,Wang:2021kxm,Shao:2022cly}. The present study provides a baseline for future investigations of these additional effects. 

The data points in the 40-60\% and 60-80\% centrality Au+Au collisions at 200 GeV~\cite{STAR:2018ldd} were used to get the blue contour in Fig.~\ref{fig:Chi2}. All available data points from both the peripheral and ultra-peripheral collisions~\cite{STAR:2018ldd,STAR:2019wlg} were used to obtain the red contour in Fig.~\ref{fig:Chi2}. The pink marker in the figure shows the result from fits to low energy electron scattering data~\cite{DEVRIES1987495}, which lies at the $3\sigma$ boundary of the red contour and gray contour and is in the middle of both in Fig.~\ref{fig:Chi2} (a) and (b). 
This indicates a possible centrality dependence. The centrality dependence may also be a potential indication of possible those four additional effects which are not included in the EPA-QED calculations. We have already demonstrated the sensitivity of the Breit-Wheeler process to small effect at the level of just a few MeV. Therefore, future high-precision measurements may potentially lead to constraint on these initial- and final-state effects. 

\section{Conclusions}
\label{sec:conclusions}
We study the collision energy dependence of the cross section and $\sqrt{\langle p_{T}^{2} \rangle}$ for electromagnetic $e^{+}e^{-}$ pair production (the Breit-Wheeler process) in heavy-ion collisions. It is found that the cross section and $\sqrt{\langle p_{T}^{2} \rangle}$ have a strong collision energy dependence. To be more specific, the cross section increases with increasing beam energy, while the $\sqrt{\langle p_{T}^{2} \rangle}$ for $e^+e^-$ pairs decreases with increasing beam energy for a fixed energy ($\omega$). Both reach a plateau above RHIC top energies for the specific kinematic acceptance in the STAR Detector discussed in this paper. We further investigate the kinematics of the pair production in order to define the criterion for the Breit-Wheeler process. It would be very interesting to test these theoretical predictions at RHIC and LHC. The collision energy dependence can be used as a powerful tool to study QED processes in strong electromagnetic fields. Moreover, the $\gamma\gamma \rightarrow l^{+}l^{-}$ process cross section and $\sqrt{\langle p_{T}^{2} \rangle}$ are sensitive to the  nuclear charge distribution in heavy-ion collisions, therefore, the nuclear radius and skin depth can be extracted by $l^{+}l^{-}$ pair $p_{T}$, $M_{ee}$ and angular distributions. Additional precision measurements at RHIC and LHC in non-UPC A+A collisions will be especially important for improved precision and sensitivity to any deviation from initial nuclear Woods-Saxon charge distribution.

\section*{Ackonwledgement}
The authors would like to thank Prof. Jian Zhou, Yajin Zhou, Cong Li, and Xin Wu for their stimulating discussion. 
This work was funded by the National Natural Science Foundation of China
under Grant Nos. 12075139, 11890713, 12175223 and 11975011, the U.S. DOE Office of Science under contract Nos. DE-SC0012704, DE-FG02-10ER41666, and DE-AC02-98CH10886. 

\bibliography{ref}

\end{document}